# Decoding of visual-related information from the human EEG using an end-to-end deep learning approach

Lingling Yang, Leanne Lai Hang Chan, and Yao Lu

Lingling Yang is with University of Minnesota. Leanne Lai Hang Chan is with Department of Electrical Engineering, City University of Hong Kong, HKSAR, China (e-mail: leanne.chan@cityu.edu.hk). Yao Lu is with the School of Data and Computer Science and Guangdong Province Key Laboratory of Computational Science, Sun Yat-Sen University, Guangzhou, China.

# Decoding of visual-related information from the human EEG using an end-to-end deep learning approach

Lingling Yang, Leanne Lai Hang Chan, and Yao Lu

*Abstract*— There is increasing interest in using deep learning approach for EEG analysis as there are still rooms for the improvement of EEG analysis in its accuracy. Convolutional long short-term (CNNLSTM) has been successfully applied in time series data with spatial structure through end-to-end learning. Here, we proposed a CNNLSTM based neural network architecture termed EEG_CNNLSTMNet for the classification of EEG signals in response to grating stimuli with different spatial frequencies. EEG_CNNLSTMNet comprises two convolutional layers and one bidirectional long short-term memory (LSTM) layer. The convolutional layers capture local temporal characteristics of the EEG signal at each channel as well as global spatial characteristics across channels, while the LSTM layer extracts long-term temporal dependency of EEG signals. Our experiment showed that EEG_CNNLSTMNet performed much better at EEG classification than a traditional machine learning approach, i.e. a support vector machine (SVM) with features. Additionally, EEG_CNNLSTMNet outperformed EEGNet, a state-of-art neural network architecture for the intra-subject case. We infer that the underperformance when using an LSTM layer in the inter-subject case is due to long-term dependency characteristics in the EEG signal that vary greatly across subjects. Moreover, the inter-subject fine-tuned classification model using very little data of the new subject achieved much higher accuracy than that trained only on the data from the other subjects. Our study suggests that the fine-tuned inter-subject model can be a potential end-to-end EEG analysis method considering both the accuracy and the required training data of the new subject.

## I. Introduction

MACHINE learning algorithms, such as support vector machine (SVM) [1] and k-nearest neighbors [2], have been applied successfully in EEG analysis. These traditional machine learning algorithms generally require the selection of discriminative features and are often applied to each subject without distinctions. Moreover, there is still significant room for improvement of EEG analysis in its accuracy.

Neural networks have become one of the most promising tools for data classification because of their ability to extract high-level discriminative features from raw low-level data. They have been developing rapidly in recent years for applications such as handwritten character recognition [3], speech recognition [4], natural language processing [5], and recommendation systems [6]. Thus, as a fully end-to-end analysis approach, neural networks hold the potential to overcome the limitations of manually selected features. Applications to electrophysiological signal [7], [8] have shown that neural networks provide much better performance than classical machine learning algorithms.

Systematic reviews of the literature [9], [10] on deep learning applications to EEG analysis reveals that convolutional neural network (CNN) and recurrent neural networks are the most efficient networks for EEG classification tasks. In this paper, we propose a convolutional long short-term (CNNLSTM) based neural network model, termed EEG_CNNLSTMNet, that classifies temporal EEG data recorded from the occipital lobe based on the spatial frequency content of the presented visual stimuli. CNNLSTM takes advantage of the complementarity of CNN for modeling the spatial input of EEG and long short-term memory (LSTM) for modeling the long-term temporal dependency of EEG signals. We compare the performance of EEG_CNNLSTMNet to a traditional machine learning algorithm and to EEGNet, a previously published neural network model [7].

EEG classification is commonly performed for the intra-subject case [11]–[13]. Moreover, the inter-subject case is becoming a matter of focused research in BCI communities, as it facilitates establishing common neural mechanisms across subjects and allows the recurring use of publicly available subjects' EEG data in brain computer interface (BCI) training [14], [15]. Thus, we evaluate our classification model for both the intra-subject and inter-subject cases.

## II. Materials

### A. Experimental Protocol

Ten volunteers (mean age: 23 ± 5 years, 2 females and 8 males) participated in the experiment in the Department of Electrical Engineering, City University of Hong Kong. All subjects were healthy with no known history of neurological disorder and had normal or corrected to normal vision. The experimental protocol was approved by the Institutional Review Board of City University of Hong Kong and was in accordance with the Declaration of Helsinki. After a detailed explanation of the experiment procedures, all subjects signed a written informed consent statement.

Subjects were seated comfortably in a chair inside a dark room and presented with a series of sinusoidal gratings on a 23-inch LCD monitor, with the monitor placed 1-meter away from the horizontal plane of the eyes. Gratings with varying spatial, temporal frequencies, and orientations, were generated using Matlab's Psychtoolbox [16]. The orientations of gratings were either horizontal or vertical. The spatial frequency was represented as the width of white and black bars, and the temporal frequency was represented as the rolling speed of the gratings on the monitor. The parameters for spatial frequencies were 0.05, 0.1, and 0.3 cycles per degree (cpd). The temporal frequencies and orientation parameters are not within the scope of the present study. Full details on the design of the experiments are given in [17].

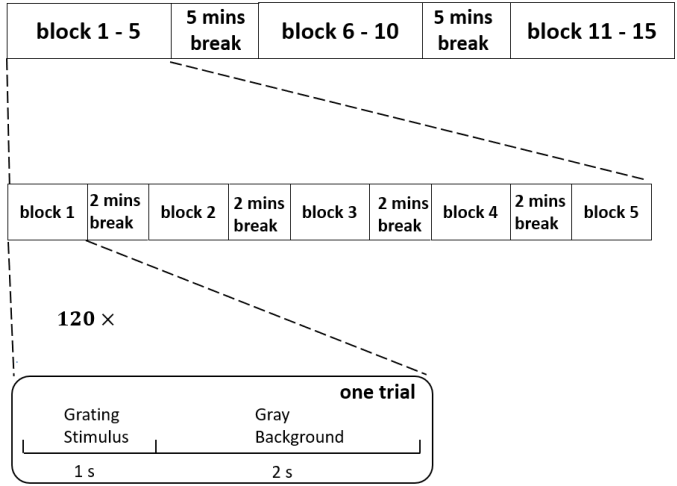

Fig. 1 Timeline of the experimental protocol

The timeline of our experiment is shown in Fig. 1. Each trial consisted of presenting a 1 s grating stimulus followed by 2 s of gray background. There were a total of 15 blocks with 120 trials each. To prevent fatigue, a time window of 2 min between consecutive blocks was included for subjects to relax (Fig. 1). Moreover, subjects were asked to relax for 5 min after every five blocks. The relaxation time could be extended if needed. For each subject the total time spent in the experiment was approximately 2 h including the relaxation time. During the experiment, subjects were instructed to fixate on the center of the monitor where a cross mark appeared as a cue. Subjects were asked to avoid blinking or any eye movements after stimulus onset.

### B. EEG Data Collection and Preprocessing

EEG signals were recorded throughout the experiment at 2048 Hz with a 34-channel active electrode system (Biosemi, Inc). The EEG acquisition system consisted of 32 scalp EEG electrodes and two reference electrodes. The EEG data were preprocessed using the open source Matlab toolbox EEGLAB [18]. The continuous EEG signals were band-pass filtered in the frequency range of 1–100 Hz (Hamming windowed sinc FIR filter), band-stop filtered at 45–55 Hz (Hamming windowed sinc FIR filter) and, down-sampled to 512 Hz. A common average reference was used successively. Then, the signals were cut into 1000 ms segments, [-200 800] ms time-locked to the onset of the gratings. To reject contaminated epochs, tests of excessive EEG values, standard deviations for single channel and for all 32 channels were performed. On average 1344 ± 247 (mean ± standard deviation) trials were analyzed per subject.

### C. Temporal and Frequency Analysis

In our previous study [17], two positive event related potentials (the P1 and P2 components) were clearly observable over the occipital lobe after the onset of the gratings. In this study, the P1 and P2 components were defined as the EEG traces during the time windows of [100 200] ms and [200 350] ms post-stimulus. And the mean amplitudes of the P1 and P2 components showed significant differences for stimuli with different spatial frequencies. Moreover, responses in the alpha and gamma bands differed for low and high spatial frequency stimuli [19]. These statistical results motivated the use of EEG data as training data for a classification model.

### III. METHODS

Here, we first introduce two classification cases. The structure of our proposed EEG_CNNLSTMNet and its input data are discussed subsequently.

### A. Classification Cases

The EEG data in response to different visual stimuli are classified for two different cases: *the intra-subject case*, and *inter-subject case*.

(1) *The intra-subject case*: In this classification case, one classification model is trained on the data of only one subject, resulting in a total of ten classification models (one per subject) (Fig. 2A). To assess classification accuracy, nested cross validation is applied to each model. On average 1210 ± 223 (mean ± standard deviation) trials were used as the training set and 134 ± 25 (mean ± standard deviation) as a test set. Several studies demonstrated that in this case, classification results for each subject vary considerably [15] – [17].

(2) *The inter-subject case*: (a) The base inter-subject case: In this classification case, the EEG data for each subject is classified by a classification model trained on the data from all other subjects (Fig. 2B). In this study, the process was repeated ten times, with the data of each subject used exactly once as the test set. Hence, our training sets contained an average of 12293 ± 363 (mean ± standard deviation) trials. This case is designed to test the feasibility of using a classification model trained on data from other subjects. (b) The fine-tuned inter-subject case: In this case, the classification model trained on the data from all other subjects is fine-tuned by adding a very small amount of data from the test subject. This case is designed to balance the trade-offs between classification accuracy and the amount of training data required for a new subject.

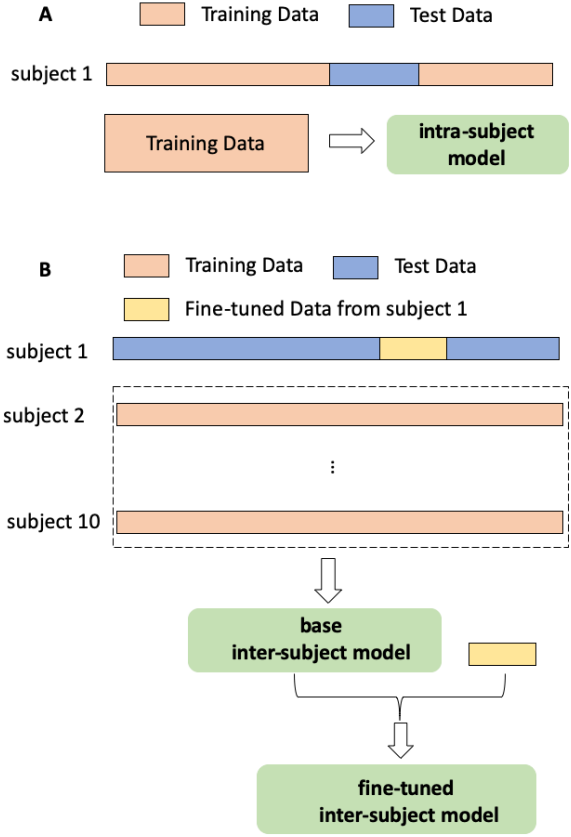

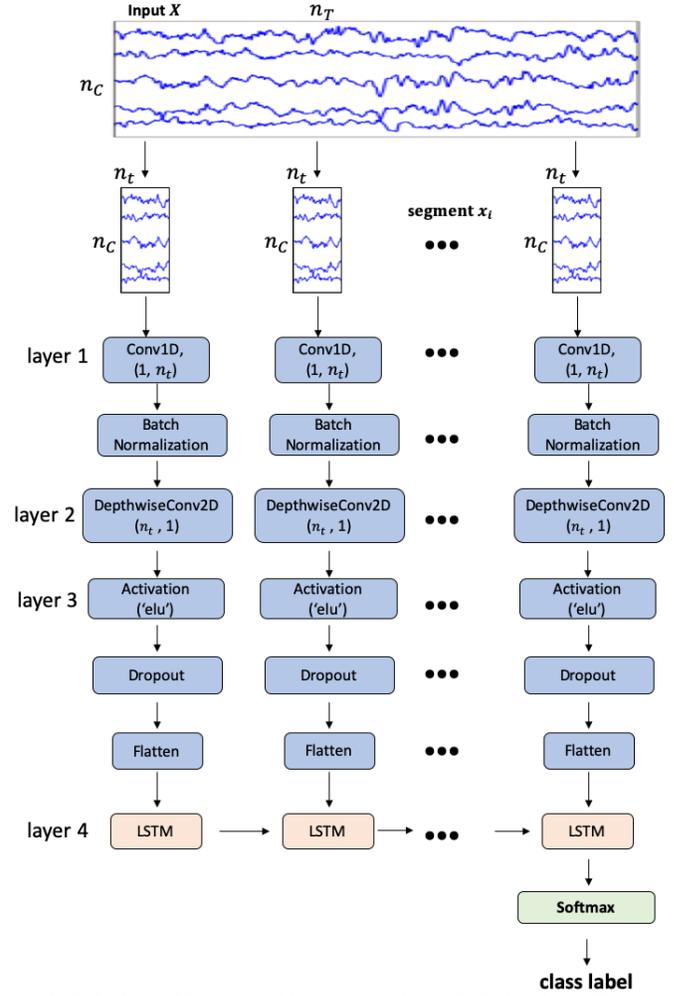

Fig. 2 Conceptual diagram for intra-subject (A) and inter-subject cases (B) of subject 1. In the intra-subject case, one intra-subject model is constructed using the training data only from subject 1. In the inter-subject case, one base inter-subject model trained on the data from subjects 2 to 10 and one fined-tuned inter-subject model through fine tuning the base inter-subject model by add a small amount of data from subject 1.

### B. EEG_CNNLSTMNet

The EEG data is time series data containing spatial information recorded at multiple sites on the scalp. Based on the spatial and temporal characteristic of the EEG data, a neural network termed EEG_CNNLSTMNet is proposed to classify EEG signal for different spatial visual stimuli. EEG_CNNLSTMNet is based on the convolutional, long short-term memory fully connected deep neural networks on vocabulary tasks [20]. CNN is good at modelling the problem related to the spatial inputs while LSTM is good at modelling the task related to the sequences. Thus, the CNNLSTM network is suitable for sequence prediction with spatial inputs, like EEG data.

Fig. 3 EEG_CNNLSTMNet architecture. The input EEG signals are $\mathbf{X} \in R^{n_C * n_T}$, where $n_C$ is the number of the EEG channels/electrodes, $n_T$ is the number of temporal samples of one trial. $x_i \in R^{n_C * n_t}$ is one segmented time interval data.

The architecture of our EEG_CNNLSTMNet is shown in Fig. 3. First, the input EEG signals $\mathbf{X} \in R^{n_C * n_T}$, ($n_C$ is the number of the EEG channels/electrodes, $n_T$ is the number of temporal samples) are segmented into non-overlapped time series $x_i \in R^{n_C * n_T}$. The segmentation process is used to avoid high frequency signals that are not useful for EEG data and to reduce the number of parameters in subsequent LSTMs. Then a short temporal convolutional filter and a spatial depth-wise convolutional followed by an exponential linear unit ('elu') activation function are used to model the spatial inputs in the segments. Then, at each time $i$, the spatial features of the current input after the flatten step and the output of the previous LSTM layer are passed together through the current LSTM cell to model the temporal structure. Finally, the representations of the input data are passed to a softmax classification to extract the classification label. To increase classification performance, the techniques of batch normalization (BN) and dropout are applied in our EEG_CNNLSTMNet. The details of the model are as follows:

*1) Convolutional Long Short-Term Memory*

In layer 1, structured 1D convolutional filters of size $(1, n_t)$ are applied to capture the short temporal characteristics of EEG signals at each channel. In layer 2, a group of spatial filters of size $(n_c, 1)$, specially designed for the EEG signals, are applied. The structural characteristics of spatial information across channels is modeled as data of one channel. Then, an 'elu' activation function is applied.

Long short-term memory is designed to avoid the issue of long-term dependency and has been applied with great success in time series analyses, such as speech recognition [21], handwriting recognition [22] and machine translation [23]. In layer 4, a hidden layer with bidirectional LSTM blocks was added to deal with the one-dimensional time series resulting from the steps of spatial convolutions.

The standard LSTM architecture including a forget gate, an input gate and an output gate was employed. At each time $t$, the LSTM cell takes in three inputs, i.e. a hidden state $\boldsymbol{h}_{t-1}$, a cell state $\boldsymbol{c}_{t-1}$ from the previous layer and the current observation $\boldsymbol{x}_t$ and gives a new hidden state $\boldsymbol{h}_t$, and a new cell state $\boldsymbol{c}_t$. The equations of one LSTM cell are as follows:

$$\boldsymbol{c}_t = \boldsymbol{f}_t * \boldsymbol{c}_{t-1} + \boldsymbol{i}_t * \widetilde{\boldsymbol{c}}_t$$
$$\boldsymbol{h}_t = \boldsymbol{o}_t * \tanh(\boldsymbol{c}_t)$$

Where $\boldsymbol{f}_t$, $\boldsymbol{i}_t$ and $\boldsymbol{o}_t$ are the forget gate, the input gate and the output gate respectively while $\widetilde{\boldsymbol{c}}_t$ is the input modulation gate. The corresponding equations of $\boldsymbol{f}_t$, $\boldsymbol{i}_t$, $\boldsymbol{o}_t$ and $\widetilde{\boldsymbol{c}}_t$ are:

$$\boldsymbol{f}_t = \sigma(\boldsymbol{W}_f[\boldsymbol{h}_{t-1}, \boldsymbol{x}_t] + \boldsymbol{b}_f)$$
$$\boldsymbol{i}_t = \sigma(\boldsymbol{W}_i[\boldsymbol{h}_{t-1}, \boldsymbol{x}_t] + \boldsymbol{b}_i)$$
$$\boldsymbol{o}_t = \sigma(\boldsymbol{W}_o[\boldsymbol{h}_{t-1}, \boldsymbol{x}_t] + \boldsymbol{b}_o)$$
$$\widetilde{\boldsymbol{c}}_t = \tanh(\boldsymbol{W}_c[\boldsymbol{h}_{t-1}, \boldsymbol{x}_t] + \boldsymbol{b}_c)$$

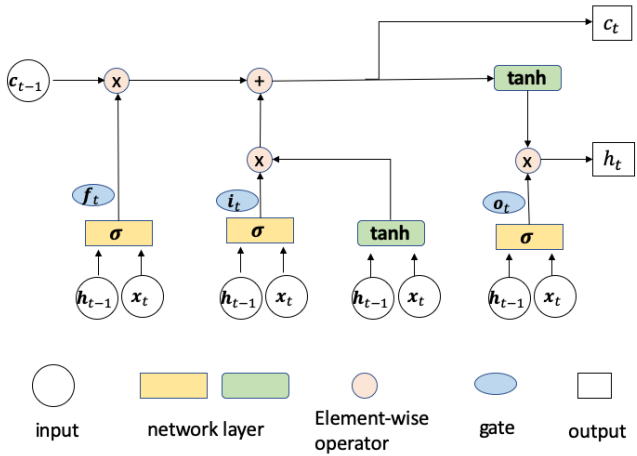

Fig. 4 The architecture of one LSTM module (Modified from [24])

*3) Batch Normalization*

As the network learns by updating its weights, the statistical distribution of its input data changes. As a result, the hidden layers keep trying to adapt, slowing down the convergence. Batch normalization on the inputs to each hidden layer ensures that the distribution remains unchanged during training [25]. The procedure of BN is as follows:

Suppose $B = \{x_{1...m}\}$ is the training set over a mini-batch, $\gamma, \beta$ are the learned parameters. The values over a mini-batch are first normalized, and then adjusted through $\gamma$ and $\beta$. The output $y_i$ of BN for each $x_i$ is calculated by:

$$\mu_B = \frac{1}{m}\sum_{i=1}^{m} x_i, \sigma_B^2 = \frac{1}{m}\sum_{i=1}^{m}(x_i - \mu_B)^2$$
$$\hat{x}_i = \frac{x_i - \mu_B}{\sqrt{\sigma_B^2 + \epsilon}}, y_i = \gamma \hat{x}_i + \beta$$

*4) Optimization*

We fit our model using the Adam optimization method. The initial learning rate is set to 0.1. When performance over the validation set decreases over two consecutive epochs, the learning rate is adjusted to 0.01. The model is trained on an NVIDIA GTX2070 GPU using the Keras API [26] with the Tensorflow backend [27]. The epoch number is set to 100 as the accuracy increment would be trivial with an epoch number higher than 100.

*5) Input Data of EEG_CNNLSTMNet*

As the objective of our study was to construct an end-to-end EEG classification model for visual stimuli, the algorithm was expected to extract the discriminative features by themselves. Based on the previous student t-test statistical analysis, we used the preprocessed temporal EEG data of P3, P4, P7, P8, PO3, PO4, O1, Oz, and O2 channels in [-200 800] ms, i.e. $R^{9*256}$.

### C. Compared Classification Models

We compared the performance of EEG_CNNLSTMNet to that of EEGNet (a state-of-the-art EEG neural network) as well as to a classical machine learning approach: support vector machine with frequency features. The power values of the alpha and gamma band of the EEG data of the P3, P4, P7, P8, PO3, PO4, O1, Oz, and O2 channels were used as frequency features.

### D. Statistical Analysis

We analyzed the statistical differences between our proposed EEG_CNNLSTMNet and the compared classification models (EEGNet, SVM with features) with a nonparametric Wilcoxon signed rank test. Throughout this study, the significance level was set to 0.05.

## IV. RESULTS

We compared the EEG classification accuracies of EEG_CNNLSTMNet in the intra-subject, and inter-subject cases to those of EEGNet and SVM with frequency features.

### A. Intra-subject Results

Classification performances with EEG_CNNLSTMNet was assessed for each subject (Fig. 5). It was showed that the classification accuracy of EEG_CNNLSTMNet was higher than SVM with features (60.08% ± 10.63% vs 45.56% ± 8.95%, $p$ = 0.012) and EEGNet (60.08% ± 10.63% vs 54.83% ± 7.77%, $p$ = 0.028). Our results also showed that EEGNet achieved better prediction performance than SVM (54.83% ± 7.77% vs 45.56% ± 8.95%, p = 0.028).

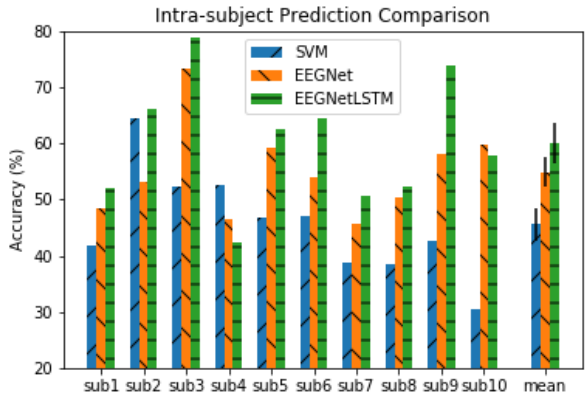

Fig. 5 Classification accuracies using SVM with frequency features, EEGNet and EEG_CNNLSTMNet in the intra-subject case. The black short vertical lines represent the standard error of the mean for SVM, EEGNet, and EEGNetLSTM respectively.

### B. Inter-subject Results

Classification results in the inter-subject case using SVM with frequency features, EEGNet, and EEG_CNNLSTMNet were shown in Fig. 6. The classification accuracies of EEG_CNNLSTMNet were higher than SVM (40.50% ± 6.22% vs 35.70% ± 6.35%, $p$ = 0.028). However, there were no statistical differences between EEG_CNNLSTMNet and EEGNet (40.50% ± 6.22% vs 38.62% ± 4.93%, p = 0.24).

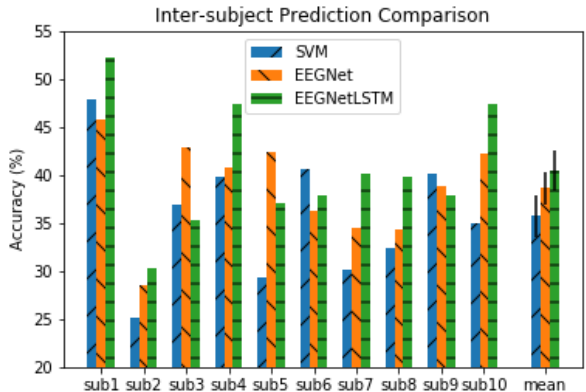

Fig. 6 Classification accuracies using SVM with frequency features, EEGNet and EEG_CNNLSTMNet in the inter-subject case. The black short vertical lines represent the standard error of the mean for SVM, EEGNet, and EEGNetLSTM respectively.

The results showed that the classification performance in the inter-subject case was much poorer than that in the intra-subject case (40.50% ± 6.22% vs 60.08% ± 10.64%, p = 0.013) (Fig. 7). The underlying reason is the distribution of EEG data across subjects may vary a lot, thus the classification model trained on the data of different subjects lacks the subject-specific information. Based on this assumption, the inter-subject classification model was fine-tuned using a very small amount (20%) of EEG data of the test subject. The classification performance of the fine-tuned classification model was increased greatly compared to the base inter-subject classification model (59.12% ± 6.56% vs 40.50% ± 6.22%, p = 0.006) (Fig. 7). In addition, there were no statistical differences between the classification accuracy of the fine-tuned inter-subject model and that of the intra-subject model (59.12% ± 6.56% vs 60.08% ± 10.64%, p = 0.575) (Fig. 7).

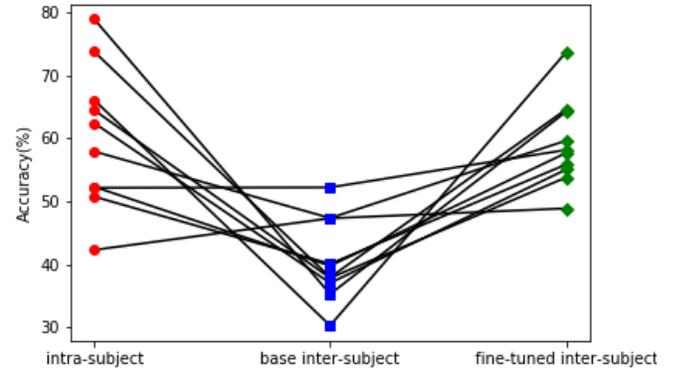

Fig. 7 Classification accuracies using EEG_CNNLSTMNet in the intra-subject, the base inter-subject and the fine-tuned inter-subject cases for all the subjects. Each red circle, blue square and green diamond represent the classification accuracies of each subject in the intra-subject, inter-subject and fine-tuned cases separately. The three points connected with a dark line belong to the same subject.

The validation accuracies along epochs during the base and the fine-tuned inter-subject model training was shown in Fig. 8. The validation accuracies can only reach about 60% during the training of the base inter-subject classification model. When a small number of data of the test subject were added, the validation accuracies decreased at the beginning and then gradually increased to a higher accuracy.

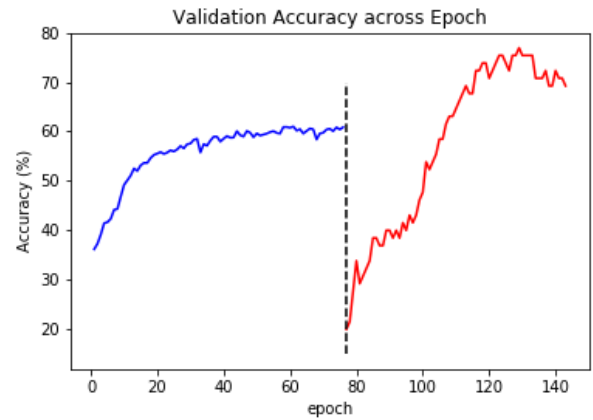

Fig. 8 Validation accuracies along epochs during the training of the base and fine-tuned inter-subject models. The blue line represents the validation accuracies during the training of the base inter-subject model while the red line represents the validation accuracies during the training of the fine-tuned inter-subject model.

The classification accuracies using the base and fine-tuned inter-subject EEG_CNNLSTMNet classification models with different rate of EEG data of the test subject were shown in Fig. 9. It clearly showed that the classification performance of the inter-subject model was improved despite of the amount of data of the test subject were used. It also demonstrated only a small portion, as low as 10%, of data from the test subject can lead to a much better performance.

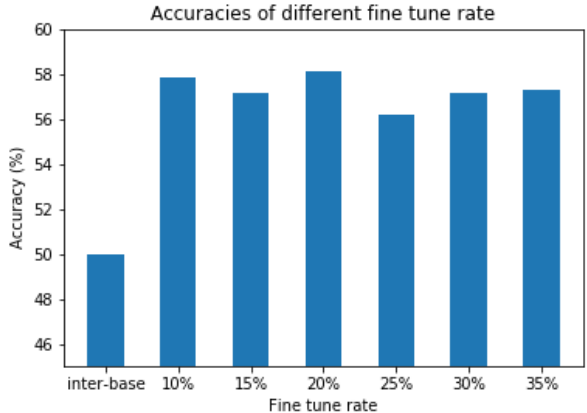

Fig. 9 Classification accuracies using the base and fine-tuned inter-subject EEG_CNNLSTMNet classification models with different rates of EEG data of the test subject.

## V. DISCUSSION

We propose a CNNLSTM based end-to-end EEG classification model for cortical response to visual stimuli, termed EEG_CNNLSTMNet, as the CNNLSTM architecture is suitable for time series data with spatial structure. In details, the first two convolutional layers sequentially model the local temporal characteristics of EEG signals at each channel and the global spatial structures across all channels. The LSTM layer models the long-term dependencies in the EEG signals. The input data of the classification model was motivated by known statistical differences in cortical responses to gratings with different spatial frequencies at different channels, i.e the P3, P4, P7, P8, PO3, PO4, O1, Oz, and O2 channels in [-200 800] ms.

We studied the performance of EEG_CNNLSTMNet in both intra-subject [15]–[17] and inter-subject cases [18], [19] – both conditions are widely studied in the neuroscience and BCI communities.

In the intra-subject case, EEG_CNNLSTMNet had significantly better performance than SVM with features and EEGNet. These results demonstrate that the first three layers of EEG_CNNLSTMNet extract discriminative features in an efficient manner. Moreover, classification based on these extracted features was superior to classification with the manually selected features. The main difference between EEG_CNNLSTMNet and EEGNet is that an LSTM layer was added to the EEG_CNNLSTMNet model. This architecture has two distinct advantages. First, LSTM can avoid the issue of long-term dependency and thus improve the model performance. And second, the size of the temporal convolution kernel can be kept small as it is unnecessary for the temporal convolutions to take into account the long-term temporal characteristics. As a result, the number of the parameters for a network with temporal convolutions followed by an LSTM layer is smaller than without an LSTM layer.

Comparisons between neural network based classifiers and traditional machine learning algorithms for the intra-subject case have been reported previously in [28]. For instance, Thomas et al. [28] demonstrated that for the classification of steady static visual evoked potential (SSVEP), convolutional neural network model is superior to SVM with features. Kwak et al. [29] also reported that SSVEP decoding results for the CNN architecture surpass those of a canonical correlation analysis (CCA)-based classifier and a CCA-$k$NN classifier. Wang et.al [30] demonstrated the ability of LSTM to shape the time-varying characteristics of EEG signals in motor imagery task. These results imply that neural networks, especially the convolutional neural network and long short-term memory network work, shows great potential in EEG signal analysis.

For the inter-subject case, we first investigated a base inter-subject classification model trained on the data from all other subjects. Our proposed EEG_CNNLSTMNet shows significantly better performance compared to the traditional machine learning method, SVM with features. Our end-to-end classification models can still obtain an accuracy of about 41%, whereas the SVM with manually selected features performs just a little above chance level (33%). This means that temporal EEG data must contain discriminative information that was not captured by the selected features. Also, the manually selected features contain subject-dependent information that affects the classification accuracy eminently. In contrast, the performances of EEG_CNNLSTMNet and EEGNet are similar in the inter-subject case. From this we conclude that the long-term dependency characteristics vary greatly across subjects, resulting in an underperformance using the LSTM layer. This is often the case in EEG signals. Based on this assumption, we demonstrated that the inter-subject classification can be fined-tuned through adding a small amount of training data from the test subject. Our results show that the fine-tuned classification model can achieve comparable classification accuracy to the intra-subject model. This suggests that the fine-tuned inter-subject model learns the subject-specific through the small amount of training data. In studies with the patients for which training set is not always easily assessable, the fined-tuned inter-subject classification model can be a potential end-to-end EEG analysis method considering both the accuracy and the amount of training data required from a new subject.

Overall, our results demonstrate the classification accuracy of the base classification in the inter-subject case was significantly lower than in the intra-subject case while the classification of the fine-tuned inter-subject model was comparable to that in the intra-subject case. This implies that it is much more challenging to classify EEG data from one subject using a single classification model trained on EEG from other subjects. The subtle variation of the distribution of EEG data for different subjects might be the underlying reason. In the intra-subject case, all the training and test data come from the same subject. So, the training and testing data come from the same underlying distribution. In the base inter-subject case, despite there are more training data due to the combination of all the data from multiple subjects, the base inter-subject model lacks the subject-specific distribution information. However, the fine-tuned inter-subject model has as much training data as in the base inter-subject case and also contains the subject-specific distribution information. Hence, compared to the intra-subject case and the base inter-subject case, the fined-tuned inter-subject model can achieve higher classification accuracy and require little training data of the new subject, which has broad application prospects for EEG.

VI. CONCLUSION

In conclusion, our proposed EEG_CNNLSTMNet classification model is more adaptive compared to feature-based machine learning algorithms. Moreover, EEG_CNNLSTMNet outperforms EEGNet in the intra-subject case. Though varying long-term dependencies across subjects reduces the performance of the base inter-subject classification models, the performance of the fine-tuned inter-subject classification model by adding a small amount of data of the new subject is improved greatly. The fine-tuned inter-subject classification model requires further investigation.

ACKNOWLEDGMENTS

This work is supported by the National Natural Science Foundation of China under grant number 11401601, by the Ministry of Science and Technology of the People's Republic of China under grant number 2016YFB0200602, by the Innovation Key Fund of Guangdong Province under grant numbers 2016B030307003, 2015B010110003, and 2015B020233008, by the Innovation Key Fund of Guangzhou grant number 201604020003, by the Natural Science Foundation of Guangdong Province under grant number 2017A030310319 and the Research Grants Council of the Hong Kong Special Administrative Region, China (Project No. CityU 9042648 and 9042829).

REFERENCES

[1] Lingling Yang, Howard Leung, David A Peterson, Terrence J Sejnowski, and Howard Poizner, "Toward a semi-self-paced EEG brain computer interface: decoding initiation state from non-initiation state in dedicated time slots," *PloS One*, vol. 9, no. 2, p. e88915, 2014.
[2] F Lotte, M Congedo, A Lécuyer, F Lamarche, and B Arnaldi, "A review of classification algorithms for EEG-based brain–computer interfaces," *J. Neural Eng.*, vol. 4, no. 2, p. R1, 2007.
[3] Ehsan Mohebi and Adil Bagirov, "A convolutional recursive modified Self Organizing Map for handwritten digits recognition," *Neural Netw.*, vol. 60, pp. 104–118, Dec. 2014.
[4] T. N. Sainath *et al.*, "Deep Convolutional Neural Networks for Large-scale Speech Tasks," *Neural Netw.*, vol. 64, pp. 39–48, Apr. 2015.
[5] Ronan Collobert and Jason Weston, "A unified architecture for natural language processing: Deep neural networks with multitask learning," presented at the Proceedings of the 25th international conference on Machine learning. ACM, 2008.
[6] A. Van den Oord, S. Dieleman, and B. Schrauwen, "Deep content-based music recommendation," in *Advances in neural information processing systems*, 2013, pp. 2643–2651.
[7] V. J. Lawhern, A. J. Solon, N. R. Waytowich, S. M. Gordon, C. P. Hung, and B. J. Lance, "EEGNet: A compact convolutional network for EEG-based brain-computer interfaces," *ArXiv Prepr. ArXiv161108024*, 2016.
[8] X. Zhai, B. Jelfs, R. H. Chan, and C. Tin, "Self-recalibrating surface EMG pattern recognition for neuroprosthesis control based on convolutional neural network," *Front. Neurosci.*, vol. 11, p. 379, 2017.
[9] A. Craik, Y. He, and J. L. Contreras-Vidal, "Deep learning for electroencephalogram (EEG) classification tasks: a review," *J. Neural Eng.*, vol. 16, no. 3, p. 031001, Apr. 2019.
[10] Y. Roy, H. Banville, I. Albuquerque, A. Gramfort, T. H. Falk, and J. Faubert, "Deep learning-based electroencephalography analysis: a systematic review," *J. Neural Eng.*, 2019.
[11] Chuang Lin, Bing-Hui Wang, Ning Jiang, Ren Xu, Natalie Mrachacz-Kersting, and Dario Farina, "Discriminative Manifold Learning Based Detection of Movement-Related Cortical Potentials," *IEEE Trans. Neural Syst. Rehabil. Eng.*, vol. 24, no. 9, Sep. 2016.
[12] Lingling Yang, Howard Leung, Markus Plank, Joe Snider, and Howard Poizner, "EEG activity during movement planning encodes upcoming peak speed and acceleration and improves the accuracy in predicting hand kinematics," *IEEE J. Biomed. Health Inform.*, vol. 19, no. 1, pp. 22–28, 2015.
[13] Fatemeh Alimardani, Reza Boostani, and Benjamin Blankertz, "Weighted spatial based geometric scheme as an efficient algorithm for analyzing single-trial EEGS to improve cue-based BCI classification," *Neural Netw.*, vol. 92, pp. 69–76, Aug. 2017.
[14] Andreas M. Ray *et al.*, "A subject-independent pattern-based Brain-Computer Interface," *Front. Behav. Neurosci.*, vol. 9, 2015.
[15] H. Kang and S. Choi, "Bayesian common spatial patterns for multi-subject EEG classification," *Neural Netw.*, vol. 57, pp. 39–50, Sep. 2014.
[16] Brainard, David H, "The Psychophysics Toolbox," *Spat. Vis.*, vol. 10, pp. 433–436, 1997.
[17] Lingling Yang and Leanne LH Chan, "An ERP study about the effects of different spatial frequencies and orientations on human brain activity," in *Engineering in Medicine and Biology Society (EMBC), 2015 37th Annual International Conference of the IEEE*, 2015, pp. 6202–6205.
[18] Delorme Arnaud and Scott Makeig, "EEGLAB: an open source toolbox for analysis of single-trial EEG dynamics including independent component analysis," *J. Neurosci. Methods*, vol. 134, no. 1, pp. 9–21, 2004.
[19] Ingo Fründ, Niko A. Busch, Ursula Körner, Jeanette Schadow, and Christoph S. Herrmann, "EEG oscillations in the gamma and alpha range respond differently to spatial frequency," *Vision Res.*, vol. 47, no. 15, pp. 2086–2098, Jul. 2007.
[20] T. N. Sainath, O. Vinyals, A. Senior, and H. Sak, "Convolutional, Long Short-Term Memory, fully connected Deep Neural Networks," in *2015 IEEE International Conference on Acoustics, Speech and Signal Processing (ICASSP)*, South Brisbane, Queensland, Australia, 2015, pp. 4580–4584.
[21] D. Amodei *et al.*, "Deep Speech 2 : End-to-End Speech Recognition in English and Mandarin," in *International Conference on Machine Learning*, 2016, pp. 173–182.
[22] A. Graves, M. Liwicki, S. Fernández, R. Bertolami, H. Bunke, and J. Schmidhuber, "A Novel Connectionist System for Unconstrained Handwriting Recognition," *IEEE Trans. Pattern Anal. Mach. Intell.*, vol. 31, no. 5, pp. 855–868, May 2009.
[23] I. Sutskever, O. Vinyals, and Q. V. Le, "Sequence to sequence learning with neural networks," in *Advances in neural information processing systems*, 2014, pp. 3104–3112.
[24] "Understanding LSTM Networks -- colah's blog." [Online]. Available: https://colah.github.io/posts/2015-08-Understanding-LSTMs/. [Accessed: 27-Oct-2019].
[25] S. Ioffe and C. Szegedy, "Batch Normalization: Accelerating Deep Network Training by Reducing Internal Covariate Shift," *ArXiv150203167 Cs*, Feb. 2015.
[26] F. Chollet, *Keras*. GitHub, 2015.
[27] M. Abadi *et al.*, "TensorFlow: Large-Scale Machine Learning on Heterogeneous Distributed Systems," *ArXiv160304467 Cs*, Mar. 2016.
[28] J. Thomas, T. Maszczyk, N. Sinha, T. Kluge, and J. Dauwels, "Deep learning-based classification for brain-computer interfaces," in *2017 IEEE International Conference on Systems, Man, and Cybernetics (SMC)*, 2017, pp. 234–239.
[29] N.-S. Kwak, K.-R. Müller, and S.-W. Lee, "A convolutional neural network for steady state visual evoked potential classification under ambulatory environment," *PLOS ONE*, vol. 12, no. 2, p. e0172578, Feb. 2017.
[30] P. Wang, A. Jiang, X. Liu, J. Shang, and L. Zhang, "LSTM-Based EEG Classification in Motor Imagery Tasks," *IEEE Trans. Neural Syst. Rehabil. Eng.*, vol. 26, no. 11, pp. 2086–2095, Nov. 2018.